# INCREASING THE ACCURACY OF SOUND VELOCITY MEASUREMENT IN A VECTOR SINGLE-BEAM ACOUSTIC CURRENT VELOCITY METER

**A.N. Grekov, N.A. Grekov, S.S. Peliushenko**

Institute of Natural and Technical Systems,
RF, Sevastopol, Lenin St., 28
*E-mail: i@angrekov.ru*

Acoustic flow velocity and direction meters from the IST series, which can be integrated into various measuring systems, are considered. The simulation of the flow profile in the measuring tube was carried out depending on the free flow velocity. A measurement method has been developed that uses the transit time of repeated acoustic signals reflected from the transducers, which makes it possible to take into account time hardware delays and a change in the length of the measuring base, which affect the determination of the speed of sound and flow, while increasing the accuracy of measuring the sound speed parameters with simultaneous measuring the flow velocity and temperatures for liquid media of different densities.

**Introduction.** The main requirements for hydrological acoustic current meters are the versatility of the device (the ability to work in sea and river conditions), as well as the continuity of measurement methods and instrument installations at hydroelectric facilities that have been formed for decades. Therefore, without taking into account these requirements, it will be impossible to correctly match the results of new measurements with archival databases due to differences in dynamic characteristics, temporal and spatial averaging.

Taking into account the above requirements, instead of hydrometric turntables manufactured in Russia, or similar ones manufactured abroad Valeport-106, -108, -308 (hydrological turntables), acoustic flow meters of the IST series were developed [1, 2], which are designed to measure the flow rate, flow direction, temperature and hydrostatic pressure (depth) of water in any watercourses from the coast, coastal structures, bridges and boats. The IST-1M vector-type device was developed and manufactured on the basis of the existing IST-1 device, the technical characteristics and scope of which were published in [3, 4]. IST-1M, in addition to measuring the speed and direction of the current, temperature and depth, has the ability to measure not only the angles of inclination of the device, but also, unlike foreign analogues, its own movements according to the readings of the accelerometers installed in them.

The developed series of acoustic devices IST can be built into measuring systems on research ships; observation ships for regular meteorological or oceanographic measurements; autonomous underwater vehicles for surveys and oceanographic research; vehicles with variable towing depth - carrying a large number of sensors, analyzers and sampling devices; moored buoys that are located on the continental shelf, devices moored on the seabed that are used to monitor processes near the seabed; drifting buoys (drifters) to correct the relative current velocity vector from the impact of the wind component; submerged buoys, which are autonomous platforms that freely drift at depth and pop up from time to time to transmit data.

The accuracy of these measuring instruments is limited by the effect on the measurement result of the time delays of signals in the transmitting and receiving paths, including delays in acoustic transducers, which significantly depend on temperature and pressure and change with time. In addition, the measurement result includes the speed of sound in the medium, which is determined with limited accuracy due to the same signal delays in the transmit-receive paths.

Due to the fact that in a single-beam acoustic current velocity meter, the speed of sound propagation in the aquatic environment is interconnected with the current velocity, it is also necessary to consider the measurement of the current velocity.

**Main part.** It is known that an ultrasonic wave carries information about the flow velocity when it propagates in a flowing liquid. Velocity can be determined by measuring the propagation time of an ultrasonic wave. The principle of ultrasonic transit time meters used today was based on the fact that the speed is uniform along the path of the ultrasonic wave. The measuring channel for the speed of sound and flow is an acoustic type channel with two piezoelectric transducers (A, B) located in the measuring tube of the device with a diameter D (Fig. 1).

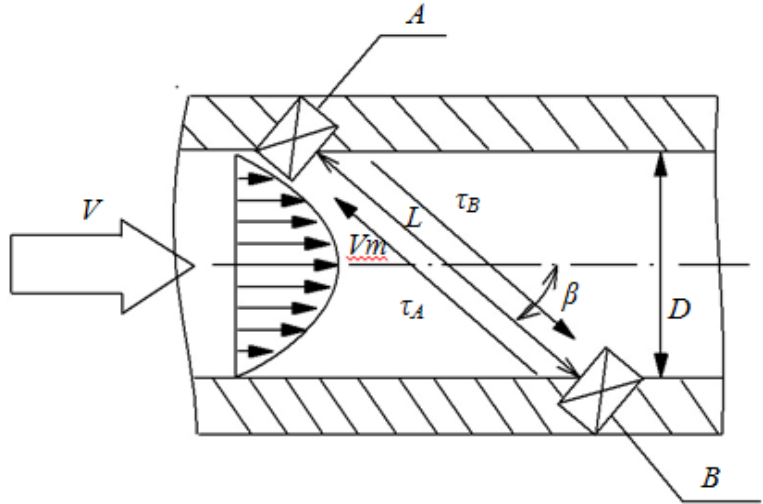

**Fig. 1.** Functional diagram of the measuring tube of an acoustic instrument

The generated electrical impulse is sequentially fed to piezoelectric transducers A and B, where it is converted into an acoustic signal that passes through the liquid and is again received by these piezoelectric transducers.
Taking into account that the piezoelectric transducers are located in a channel with a diameter D at a certain angle β, the equations for the propagation time of ultrasonic pulses in the direction of fluid movement and against it can be written as

$$\tau_A = \frac{D/\sin\beta}{C + V\cos\beta}, \quad \tau_B = \frac{D/\sin\beta}{C - V\cos\beta}, \quad (2)$$

where V is the fluid flow rate; C is the speed of sound in water.

With some assumptions, namely, that the flow velocity profile in the acoustic measuring channel is homogeneous and, excluding time delays and sound propagation velocity C, taking into account that C >> V from expression (2), finally simplify The -valued expression for determining the flow velocity can be written as

$$V = \frac{D(\tau_A - \tau_B)}{\tau_A \tau_B \sin 2\beta}. \quad (3)$$

However, it is well known that the flow velocity profile in an acoustic measuring channel is nonuniform both in laminar and turbulent flows [5–7]. An analysis of known works [8] showed that there is a limited amount of available information about the influence of the velocity profile on the measurement uncertainty. Iooss B. et al. [9] proved that the flow velocity is overestimated due to the assumption of a uniform distribution velocity, and the correction factor is obtained empirically from the results of numerical simulations for fully developed turbulence. Hui Zhang et al. [10] gave a correction factor for the transition region with a Reynolds number in the range of

2000–20000 based on the results of the Particle experiment. The correction factor for transit time has not been sufficiently explored by theoretical analysis. The article [11] theoretically and experimentally considered the influence of velocity profiles across the acoustic measuring channel on the propagation time of an ultrasonic wave. Theoretical correction factors are proposed to improve the measurement accuracy. For laminar flow, the correction factor is 0.75. For turbulent flow, the correction factor depends on the Reynolds number, and experimental results have shown that the proposed correction factors are in good agreement with the theoretical correction factors and their average relative error is defined as 0.976% for laminar flow and 0.25% for turbulent flow.

It is not possible to use these results for our case, since the known works do not take into account the influence of the confuser, the limited length of the measuring channel, and the oblique jet flow. Note that the measuring acoustic channel of IST devices is distinguished by its peculiarity, which lies in the fact that in addition to the measuring tube, which has a certain inner diameter and is tied to the dimensions of the entire device, there is also a confuser that provides low drag and good flowability in the speed range up to 5 m/s at external pressures of about 10 MPa. Taken together, the dimensions of the developed device do not exceed the dimensions of mechanical turntables, which makes it possible to use it to measure currents starting from depths of 5 cm.

The device was developed using the AutoCAD software package [12]. On fig. 2 shows a general view of the device. The diameter of the measuring tube and the opening angle of the confuser for a particular meter were determined as a result of modeling on the integration CAD and CAE systems, in particular, the COSMOSFloWorks application of the SolidWorks package [13], which were also used to study the fields of flow velocity and pressure when flowing around devices of the IST series.

One of the studies carried out in the COSMOSFloWorks application of the SolidWorks package with the IST-1 flow meter is modeling the flow profile in the measuring tube and determining the error from oblique jet in the angle range from 0° to 30° and comparing the results with errors of mechanical turntables.

The mathematical model of the computational domain was a parallelepiped with overall dimensions lxbxh = 0.5 x 0.4 x 0.4 m3.

To evaluate the accuracy of the obtained solution of the set mathematical problem, several calculations were carried out on different computational grids, differing in size and, accordingly, in the number of cells. It was found that, starting from a certain grid frequency, the solution of the problem ceased to depend significantly on the grid frequency, i.e., came out on the "shelf", which indicated that the required accuracy of the solution of the mathematical problem was achieved, since the grid convergence of the solution of the mathematical problem was obtained.

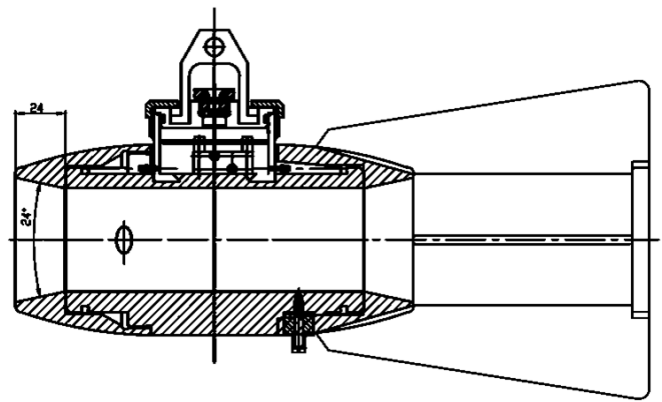

**Fig. 2.** General view of the device

Thus, modeling with the help of COSMOSFloWorks consisted in carrying out, at fixed values of the model, computational domain, boundary and initial execution conditions, several calculations with variation of the computational grid. Determination of the accuracy of the obtained solution.

On fig. Figure 3 shows the results of flow simulation in the IST-1 device at a flow velocity of 1 m/s. Then, using the obtained flow velocity profiles, the average flow velocity between two piezoemitters A and B located at a distance was determined:

$$V = \frac{1}{L}\int_0^L V(L)dL.$$

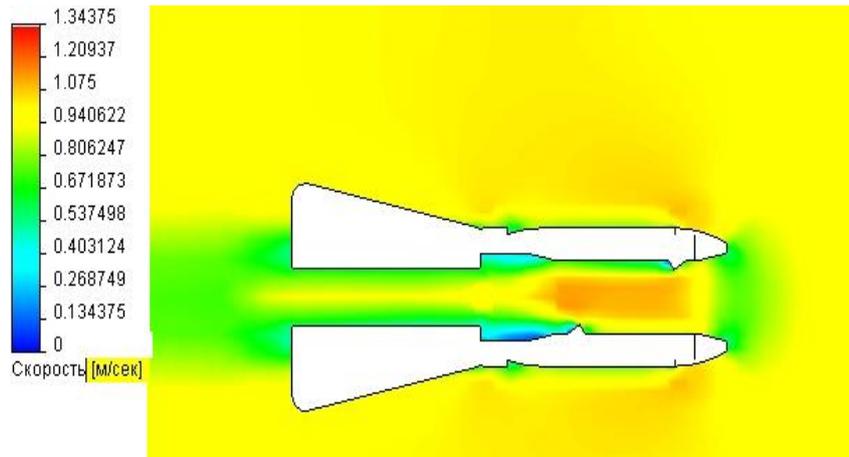

**Fig. 3.** Flow velocity field, results of flow simulation in the IST-1 device at a flow speed of 1 m/s

To analyze the simulation results, the profiles of the flow velocity in the measuring tube were constructed depending on the free flow velocity with a resolution of 1 m/s and a maximum velocity of 5 m/s. On fig. Figure 4 shows a graph of the flow velocity profile in the acoustic channel of the IST-1 instrument at a flow velocity of 2 m/s and zero skew jetting.

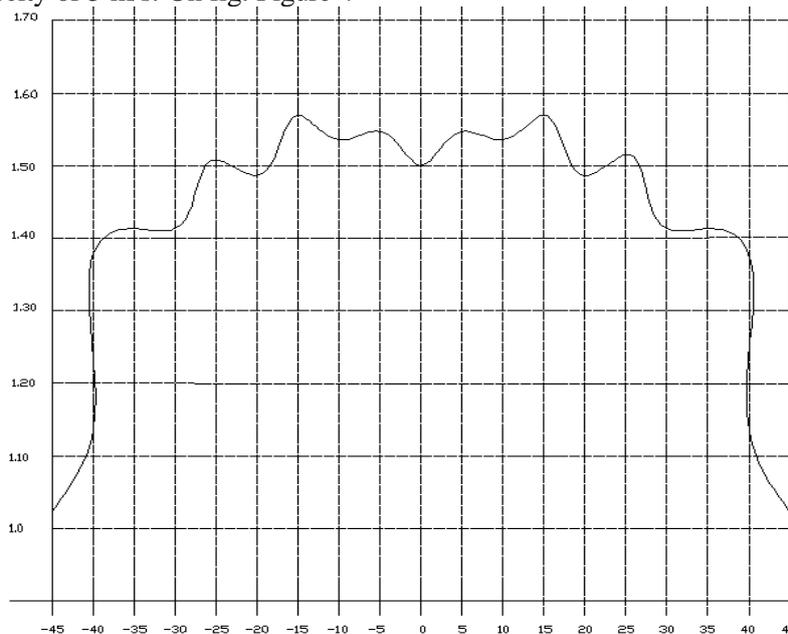

**Fig. 4.** Graph of the flow velocity profile in the acoustic channel of the IST-1 instrument at a flow rate of 2 m/s and zero skew

The non-uniformity of the flow velocity profile (see Fig. 4) in the measuring channel of the device is caused by the limited cross-section of the hydroflume, the walls of which were located at a distance of 0.15 m from each other during the simulation. With an increase in the cross section to 0.5 m, the peaks in the diagram decreased and did not appear for an open pool.

Further, by changing the position of the instrument relative to the velocity vector of the oncoming flow in the horizontal and vertical planes with a resolution of 5°, the profiles of the flow velocities in the measuring channel of the instrument were plotted. On fig. Figure 5 shows the flow velocity field, the results of flow simulation in the IST-1 device at a flow velocity of 1 m/s.

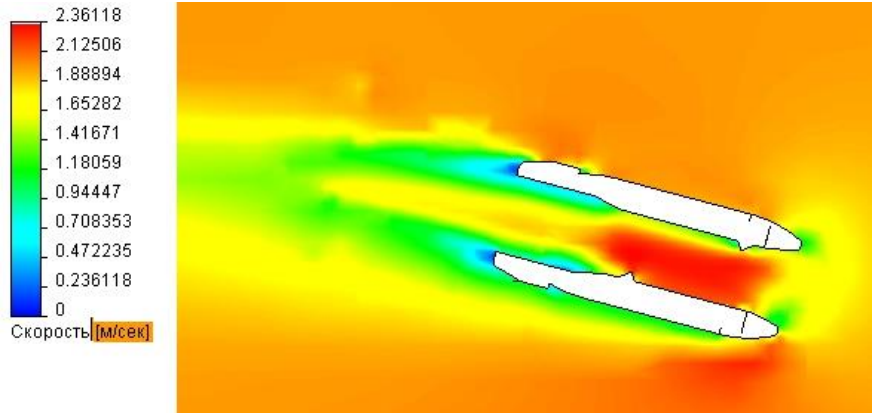

**Fig. 5.** The results of modeling the flow of velocities with the model of the IST-1 device with its inclination by 15° and flow velocity of 2 m/s

Based on the simulation results (see Fig. 5), a graph was plotted for the current velocity profile passing between two piezo-electric emitters A and B, located at a distance and inclination of the IST-1 instrument by 15° and a current velocity of 2 m/s, which presented in fig. 6.

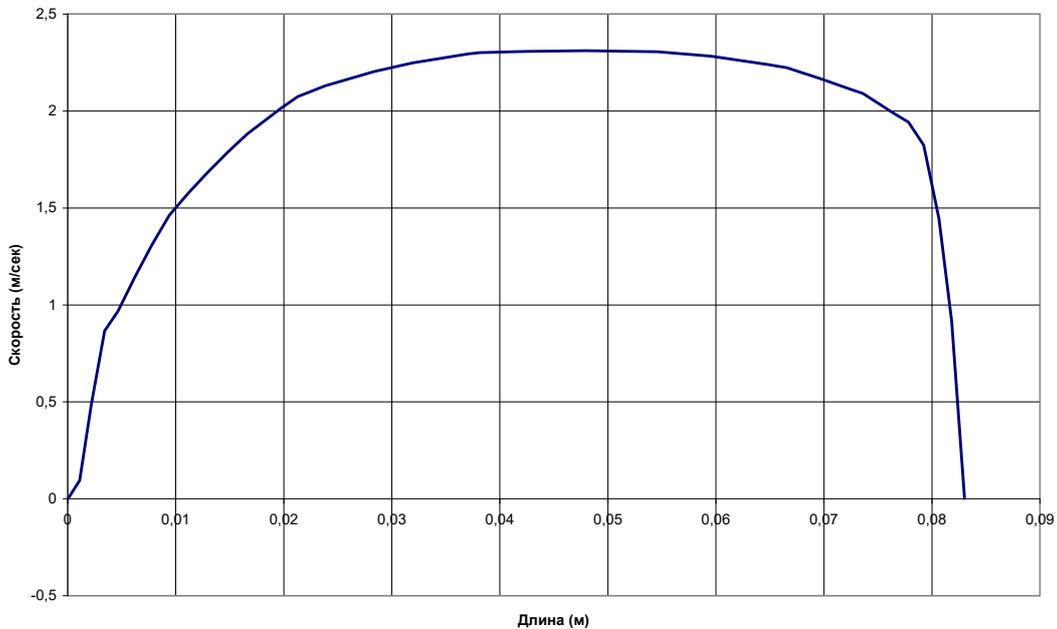

**Fig. 6.** Graph of the profile of the flow velocity passing between two piezo emitters *A* and *B*, located at a distance and inclination of the IST-1 device by 15° and a flow velocity of 2 m/s

As follows from the simulation results, the flow velocity in the profile $V = \frac{1}{L}\int_0^L V(L)dL$, $V(L)$ has a non-linear character and depends on the velocity of the oncoming flow and its oblique jet. For this study, 50 plots of free stream velocity were constructed with a resolution of 1 m/s at flow velocities of up to 5 m/s with a change in the position of the device relative to the free stream vector in the horizontal and vertical planes with a resolution of 5°.

As a result of the simulation, it was determined that with a skew jet of more than 35 °, a sharp drop in the average flow velocity between the emitters of the device relative to the value of the oncoming flow begins, which leads to significant errors in the IST-1 meter in terms of velocity. Similarly, for hydrodynamic turntables, a sharp decline begins after 30 °. All obtained modeling data were additionally confirmed as a result of full-scale tests in a hydrodynamic flume.

It follows from these studies that it is not possible to create a mathematical model for calculating the flow velocity, taking into account all the influencing factors and with acceptable accuracy. Therefore, the values $\tau_A$ and $\tau_B$ in equation (3) were determined in natural conditions according to the results of calibration carried out on a special hydrodynamic flume.

Now let's consider a method for improving the accuracy of measuring the speed of sound during the passage of acoustic signals in the measuring channel of the device, taking into account delays. The block diagram of the acoustic measuring channel with delays is shown in fig. 7.

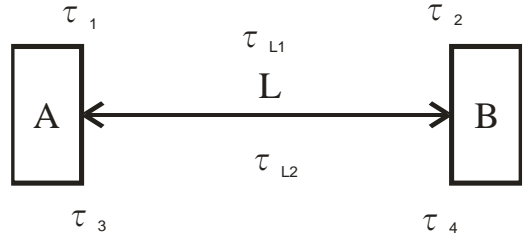

**Fig. 7.** Structural diagram of an acoustic measuring channel with delays

Here A and B are acoustic piezo transducers located in the measuring tube at a distance L.

Each of the piezoelectric transducers has a signal delay when transmitting $\tau_1$ and $\tau_3$ and when receiving - $\tau_2$ and $\tau_4$. Acoustic signal transit time from transducer A to transducer B:

$$\tau_A = \tau_1 + \tau_{L1} + \tau_2, \quad (4)$$

from transducer B to transducer A:

$$\tau_B = \tau_3 + \tau_{L2} + \tau_4, \quad (5)$$

where $\tau_{L1} = L/(C+V)$ and $\tau_{L2} = L/(C-V)$.

Additionally, when emitted by transducer A, we receive the signal reflected from transducer B by the same transducer

$$\tau_{A0} = \tau_1 + \tau_{L1} + \tau_{L2} + \tau_3. \quad (6)$$

When transducer B emits and receives the reflected signal

$$\tau_{B0} = \tau_4 + \tau_{L2} + \tau_{L1} + \tau_2. \quad (7)$$

The sum of signal times $\tau_A$ and $\tau_B$ can be written as:

$$\tau_A + \tau_B = \tau_1 + \tau_2 + \tau_3 + \tau_4 + \frac{L}{C+V} + \frac{L}{C-V}, \quad (8)$$

$$C = \frac{L \pm \sqrt{L^2 + \Delta\tau^2 V^2}}{\Delta\tau}, \quad (9)$$

where $\Delta\tau = \tau_{A0} + \tau_{B0} - \tau_A + \tau_B$ – the value of the difference signal in the measured liquid.

When the temperature T is exposed to the measuring channel, we write the equation as:

$$C(T) = \frac{L(T) \pm \sqrt{L^2(T) + \Delta\tau^2 V^2}}{\Delta\tau}. \qquad (10)$$

The length of the measuring base for a specific temperature T is determined from the equation

$$L(T) = C_\partial(T)\Delta\tau_\partial,$$

where $\Delta\tau_Д(T) = \tau_{A0} + \tau_{B0} - \tau_A - \tau_B$, $C_\partial(T)$ – temperature dependence of sound speed in distilled water. For calculations, the well-known fifth-order equation [14] of the dependence of the speed of sound on temperature with a standard deviation of 0.0028 m/s for 148 observations between 0.001 and 95.126°C on the T68 scale is used. It is considered that the accuracy of the equation is 0.015 m/s, and the reproducibility of repetitions is 0.005 m/s.

$$C_\partial(T) = 1{,}40238744 \cdot 10^3 + 5{,}03836171 \cdot T - 5{,}81172916 \cdot 10^{-2} \cdot T^2 + 3{,}34638117 \cdot 10^{-4} \cdot T^3 - 1{,}48259672 \cdot 10^{-6} \cdot T^4 + 3{,}16585020 \cdot 10^{-9} \cdot T^5.$$

Methodologically, the work consisted in the following: the IST series device was placed in a thermostat with distilled water, where the temperature T was controlled with an LTA/BE 69551-17 exemplary thermometer; the temperature was additionally controlled by an internal measuring channel located in the device. At the same time, time delays of signals were recorded at fixed temperatures, obtained from a microcontroller built into the structure of the device, then the delay values were calculated $\tau_{A0_i}$, $\tau_{B0_i}$, $\tau_{A_i}$, $\tau_{B_i}$ as a function $f(T_i)$. According to the obtained values, the dependence of the length of the measuring base L on the temperature T was determined, which was approximated by a polynomial. The graph of the increment in the length of the measuring base ΔL on temperature T is presented in fig. 8.

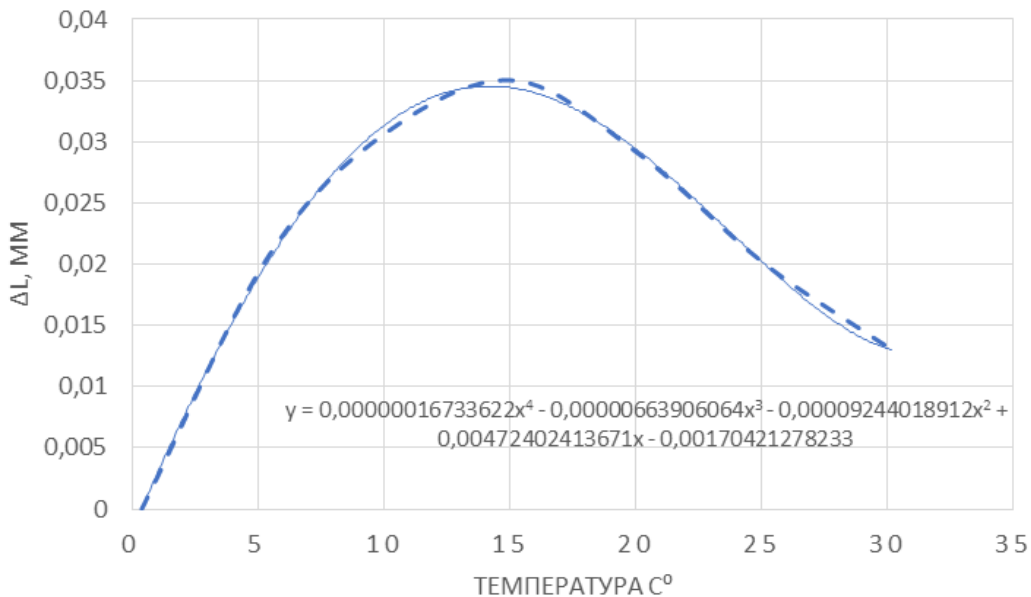

**Fig. 8.** Dependence of the length increment of the measuring base ΔL on temperature T

To automate the measurement process, the reduced curve (Fig.8) was approximated by a polynomial

$$y = 0{,}00000016733622x^4 - 0{,}00000663906064x^3 - 0{,}00009244018912x^2 + \\ + 0{,}00472402413671x - 0{,}00170421278233.$$

Let us estimate the influence of the flow velocity V on the measurement error of the sound velocity C, provided that all time delays of signals in the transmitting and receiving paths, including delays in acoustic transducers, and the change in the length of the measuring base are taken into account according to the method presented above. For the calculation we use equation (9). As a result of the calculation, a graph of the dependence of the sound speed increment ΔC on the flow velocity V was obtained, which is shown in fig. 9.

From the graph (Fig. 9) it can be seen that theoretically the maximum sound speed error does not exceed 1.6 mm/s for this instrument design in the oceanographic range of sound speed changes and in the speed range of 0–5 m/s.

Thus, the proposed method makes it possible to provide an increased accuracy of sound velocity measurement while simultaneously measuring flow velocity and temperature for liquid media with different densities.

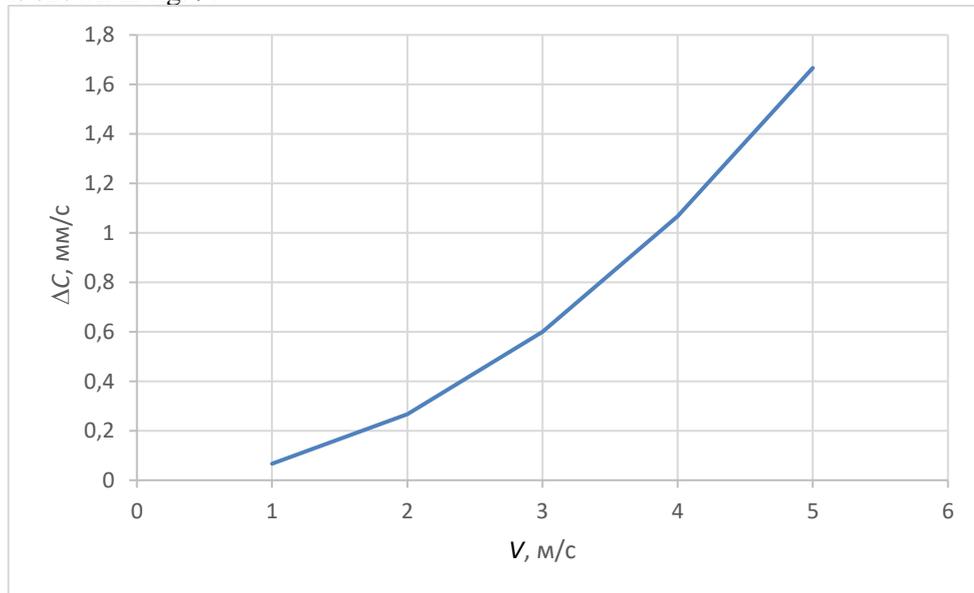

Fig. 9. Graph of the increase in the speed of sound from the speed of the current

**Conclusion.** A measurement method has been developed that uses the transit time of repeated acoustic signals reflected from the transducers, which makes it possible to take into account time hardware delays and a change in the length of the measuring base, which affect the determination of the speed of sound and flow, while increasing the accuracy of measuring parameters speed of sound with simultaneous measurement of flow velocity and temperature for liquid media with different densities.

The modeling of the flow velocity profiles in the measuring tube of a device with a confuser was carried out depending on the skew jet and oncoming flow velocity, with a resolution of 1 m/s and a maximum velocity of 5 m/s.

An analysis was made of the flow velocity profiles obtained between two piezoelectric emitters located at a distance of , from which it follows that it is not possible to create a mathematical model for calculating the flow velocity, taking into account all influencing factors and with acceptable accuracy. Therefore, the delay values must be determined in natural conditions, using the results of calibration carried out on a special hydrodynamic flume.